\documentclass[10pt,conference]{IEEEtran}

\usepackage{cite}
\usepackage{amsmath,amssymb,amsfonts}
\usepackage[ruled,vlined,linesnumbered]{algorithm2e}
\usepackage{graphicx}
\usepackage[table]{xcolor}
\usepackage{booktabs}
\usepackage{multirow}
\usepackage{url}
\usepackage{caption}
\usepackage{subcaption}
\usepackage[official]{eurosym}
\usepackage{balance}
\usepackage{soul}

\newcommand{\IGNORE}[1]{}

\newcommand{\Sip}{Dpto. de Sistemas Inform\'{a}ticos y Computaci\'{o}n} 
\newcommand{\Ucm}{Universidad Complutense de Madrid}

\definecolor{lightGreen}{RGB}{144,238,144}
\definecolor{lightRed}{RGB}{255,168,168}

\begin{document}

\title{Detection of Smart Grid Integrity Attacks Using Signal Temporal Logic}



\author{
\IEEEauthorblockN{Jos\'e Ignacio Requeno}
\IEEEauthorblockA{\textit{\Sip} \\
\textit{\Ucm}\\
Madrid, Spain \\
jrequeno@ucm.es}
}

\maketitle

\begin{abstract}
Cyber-attacks can have severe impacts on critical infrastructures, from outages to economical loss and physical damage to people and environment. One of the main targets of these attacks is the smart grid.
In this paper, we propose a new software detector for integrity attacks targeting smart meter readings. 
The detector relies upon mining parameters of temporal logic specifications for integrity attack classification. To this end, we use Signal Temporal Logic (STL) for specifying properties over time series. 
Our approach considers different ``attack scenarios" found in last years: given a parametric formula for each ``attack scenario" and a set of labeled traces, we aim at finding the parameter valuation that validates each template.
\end{abstract}

\begin{IEEEkeywords}
integrity attacks, smart grids, anomaly-based detection, data mining, temporal logic, Signal Temporal Logic 
\end{IEEEkeywords}

\graphicspath{{fig/}}


\section{Introduction}
\label{sec:intro}

Smart grids are complex cyber-physical systems (CPS), built on top of electrical power infrastructures. 
Their Advanced Metering Infrastructure (AMI) allows better maintenance performance (e.g., more timely remote diagnosis of meters issues and service connection/disconnection) and efficient monitoring while keeping consumers informed on their consumption habits.
However they are vulnerable to cyber-attacks~\cite{IET16}, which may cause 
information leaks~\cite{LMDS10}, 
energy theft~\cite{MHFBZ13} and many other security, safety or economical issues~\cite{KLBPDM19}.

In this ongoing work, we focus on the attacks that aim at manipulating the consumers' energy consumption for their benefit once they manage to hack into the AMI system~\cite{EEI2026}.
The detection and identification of frauds started with statistical techniques~\cite{KLSH04,BH02}. 
Since then, a variety of software approaches to detect malicious attacks in power systems appeared due to the fast development 
and exposition of AMI in smart grids~\cite{JLWLSS14, WCHK18}.
Usually, AMI energy-theft attacks are based on fault data injection or mimicking customers' consumption profiles.
Electricity theft detection methods range from specialized methods on well-defined attack strategies to the discovery of general consumption behavior anomalies~\cite{JAL15}.
The goal is to distinguish between normal and anomalous energy usage patterns in order to classify the samples in the dataset into one of the predefined attack profiles.

In this paper, we aim at proposing a new anomaly detection technique to identify and classify \emph{suspicious} behaviors such as data manipulation
that occurs in AMI, based on the data collected from the smart meter.

The technique relies upon mining formal specifications from observed traces (i.e., smart meter records) for the classification of attacks.
In particular, we use Signal Temporal Logic (STL) \cite{STL}, a formalism for specifying properties over time series.
In this work, we aim at developing a new attack classifier that is more robust in terms of accuracy and explainability than other (un)supervised data mining methods.

We plan to assess the performance and effectiveness of the new detector by using the evaluation framework from our previous work~\cite{EDCC21},
where  different ``attack scenarios" found in last years are considered.
 %
\section{Detection based on mining specifications} 
Specification mining looks for extracting formal specifications (i.e., automata or logical expressions) that represent a set of behaviors in a rigorous way.
Most of the works use Signal Temporal Logic (STL) \cite{STL} as the formal specification language for mining requirements in CPS \cite{KJB17, HDF18}.
Some of them propose methods to obtain the formula from scratch while other approaches start with a preliminary template (i.e., formula with parameters) and search for the parameter valuations that align with the records in the CPS.
In our case, we follow the second approach.
We address the detection of integrity attacks on smart meters as a problem of multi-class classification, where each class represents an attack scenario.

The main issues to be tackled in the construction of the prediction model are: 1) to define a template formula and 2) to find the set of parameters settings that discriminate multiple classes of attacks.
To this end, we use Parametric Signal Temporal Logic (PSTL) \cite{PSTL} to formalize the specification of templates describing the attacks.
In other words, we define each attack scenario by a parametric temporal logic specification and our goal is to mine the range of parameter valuations that characterizes an attack using a set of daily records from a smart meter. 
Existing tools such as ParetoLib \cite{ParetoLib} help computing the set of parameters settings by transforming the parameter synthesis into a multicriteria optimization problem.

\begin{figure}[htb]
	\centering
    \includegraphics[width=.75\columnwidth]{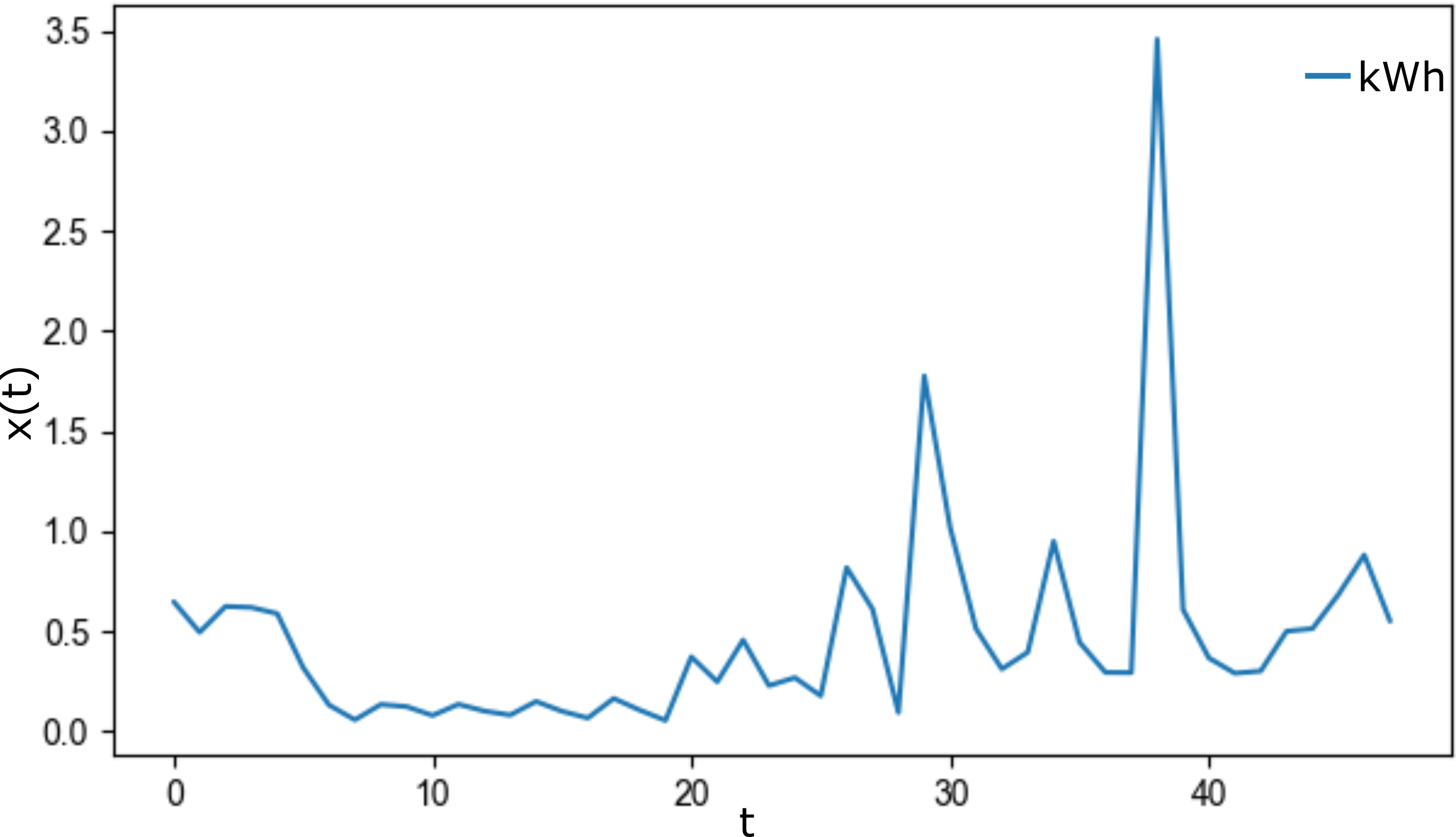}
    \caption{Power consumption for one day starting at midnight.} 
    \label{fig:day}
\end{figure}

\begin{figure}[htb]
	\centering
    \includegraphics[width=.75\columnwidth]{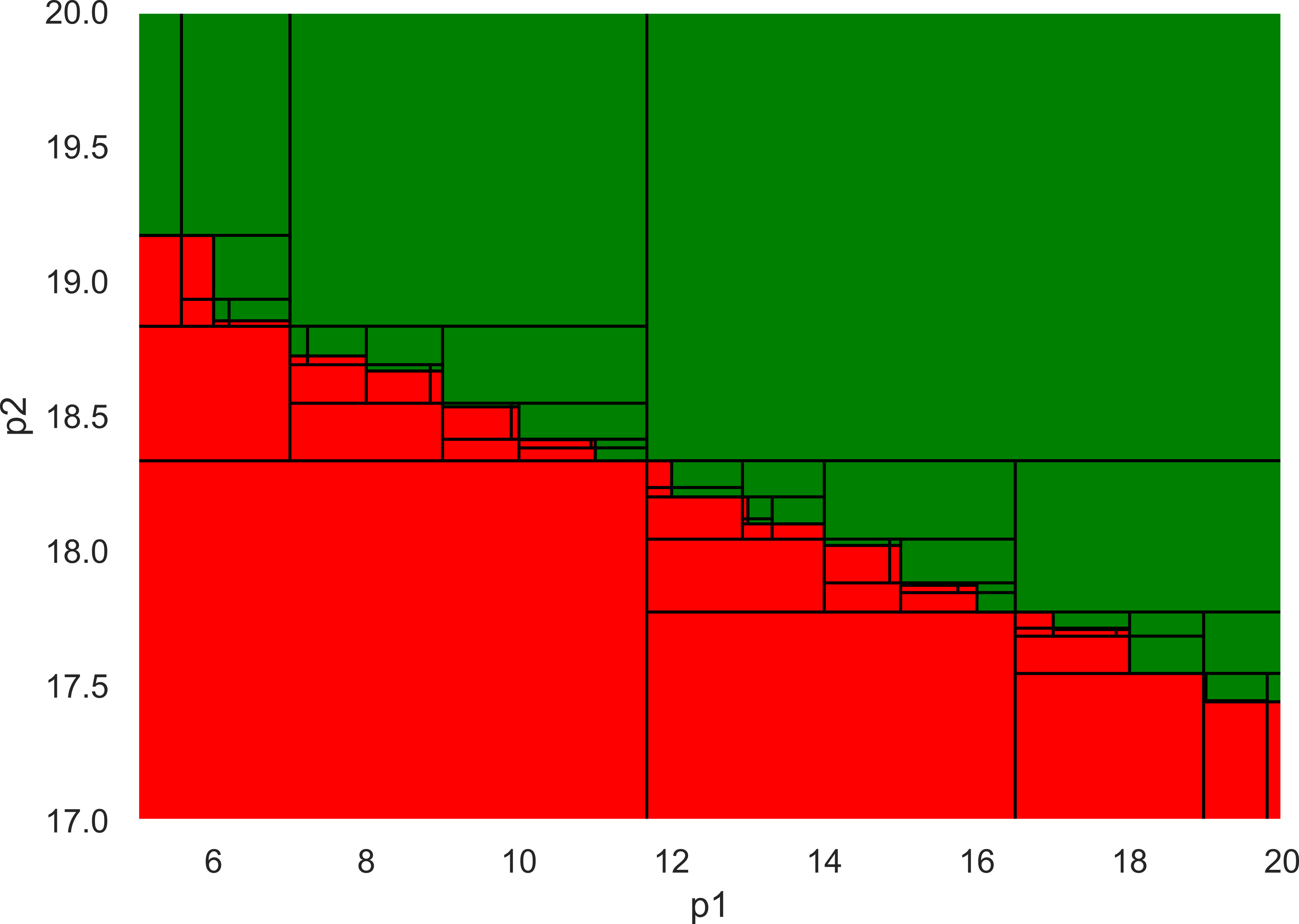}
    \caption{Combinations of p1 \& p2 with regards to Formula~\ref{eq:pstl}.}
    \label{fig:pareto}
\end{figure}

For instance, regular users usually have a higher consumption during the evening, when they are back home, and not too much during the night.
Fig. \ref{fig:day} illustrates this pattern.
It shows the power consumption recorded by a smart meter for one day (samples are taken every 30 minutes)~\cite{ISSDA-CER}. 
Fig. \ref{fig:pareto} shows whether the accumulated power consumption is below a threshold (p2) or not for the period [p1, 48] as represented by PSTL Formula~\ref{eq:pstl} with integrals (I): the green area represents the combinations of values for parameters p1 and p2 that satisfy the formula, while the red area represents the combination of p1 and p2 that don’t satisfy the formula.

\vspace{-0.5cm}
\begin{center}
\begin{equation}
On_{[p1, 48]} I \, \mathbf{x} < p2
\label{eq:pstl}
\end{equation}
\end{center}


\vspace{-0.4cm}

\vspace{-0.4cm}

Anomalous behaviors or attacks are detected if they return consumption profiles different from Fig. \ref{fig:pareto}.
We aim at proving through our experiments in our ongoing work that the integrity attack detector based on mining formal specifications will provide more robust (i.e., explainable and accurate) results than integrity attack detectors based on any other (un)supervised data mining methods.

Besides, this type of approach usually requires a few training samples to tune the parametric specification.
However, the most complex part of our proposal consists of getting the parametric temporal logic specification that clearly distinguishes each attack scenario. 
Finally, the computation of exact solutions for the parameters settings is time consuming but approximate methods exist. 
\medskip

\section{Conclusion}
\label{sec:concl}
In this paper, we show an ongoing work for the implementation of an integrity attack detector.
The integrity attack detector is based on mining temporal logic specifications from labeled traces.
Each attack is defined as a parametric temporal logic specification with exclusive validity domain for each parameter.
In particular, we propose Signal Temporal Logic as specification language.
We will apply data mining techniques for learning the validity domain of parametric temporal logic specifications.
We expect that an integrity attack detector based on parametric temporal logic specifications will provide more robust results in terms of accuracy and explainability than integrity attack detectors based on any other (un)supervised data mining methods.
In our future work, we have to complete an exhaustive experimental evaluation of our approach and compare the results with existing integrity attack detectors. 



\section*{Acknowledgment}

This work was supported by the Spanish Ministry of Science and Innovation under project 
 FAME (RTI2018-093608-B-C31) and the Comunidad de Madrid under project FORTE-CM (S2018/TCS-4314) co-funded by EIE Funds of the EU. 
\bibliographystyle{ieeetr}
\bibliography{biblio}


\end{document}